\documentclass[12pt]{iopart}

\usepackage{iopams}  
\usepackage[dvips,dvipdfm]{graphicx}

\newcommand{\gtsim}{\protect\raisebox{-0.5ex}{$\:\stackrel{\textstyle >}
	{\sim}\:$}}

\newcommand{\img}{{\rm i}}

\newcommand{\mC}{\mathbb{C}}
\newcommand{\mR}{\mathbb{R}}
\newcommand{\mN}{\mathbb{N}}

\begin{document}

\title[Analytic properties of random energy models I: moment of 
the partition function]
{On analyticity with respect to the replica number in 
random energy models I: an exact expression of the moment of 
the partition function }

\author{Kenzo Ogure$^1$ and Yoshiyuki Kabashima$^2$}

\address{$^1$
Department of Nuclear Engineering, 
Faculty of Engineering, Kyoto University, 
Kyoto 606-8501, Japan \\
$^2$
Department of Computational Intelligence and Systems Science, 
Tokyo Institute of Technology, Yokohama 226-8502, Japan\\}
\ead{ogure@nucleng.kyoto-u.ac.jp$^1$, kaba@dis.titech.ac.jp$^2$}
\begin{abstract}
We provide an exact expression of the moment of the partition 
function for random energy models of finite system size, 
generalizing an earlier expression for a grand canonical version of the discrete random energy model
presented by the authors in {\em Prog. Theor. Phys. } {\bf 111}, 661 (2004).
The expression can be handled both analytically and numerically, 
which is useful for examining how the analyticity 
of the moment with respect to the replica numbers, 
which play the role of powers of the moment, 
can be broken in the thermodynamic limit. 
A comparison with a replica method analysis
indicates that the analyticity breaking can be 
regarded as the origin of the one-step replica symmetry breaking.  
The validity of the expression is also confirmed by 
numerical methods for finite systems. 
\end{abstract}

\maketitle

\section{Introduction}
The replica method (RM) is one of the few analytical techniques 
available for dealing with disordered systems 
\cite{Beyond1987,Dotzenko2001}. 
In general, RM can be regarded as a scheme for evaluating 
the generating function 
\begin{eqnarray}
\phi_N(n)=\frac{1}{N} \log \left \langle Z^n \right \rangle, 
\label{gen_func}
\end{eqnarray}
for real number $n \in \mR \ (\mbox{or complex} \ \mC)$, 
where $Z$ and $\left \langle \cdots \right \rangle $
represent the partition function and the average over the quenched 
randomness that governs the objective system, respectively. 
Unfortunately, direct evaluation of eq. (\ref{gen_func})
is difficult in general. In statistical mechanics, 
the system size $N$ is in most cases assumed to be 
infinitely large, which often makes the analysis possible for
$n=1,2,\ldots \in \mN$. 
Therefore, RM in statistical mechanics
usually first evaluates eq. (\ref{gen_func}) for 
$n \in \mN $ in the limit $N \to \infty$ as
\begin{eqnarray}
\phi(n)=\lim_{N \to \infty} \phi_N(n),
\label{phi}
\end{eqnarray}
and then analytically continue the expression from 
$n \in \mN$ to $n \in \mR$.  

This procedure generally includes at least two mathematical problems,
although it was recently proved for several examples
that the solutions obtained by RM are mathematically 
correct \cite{Talagrand2003,Talagrand2006}. 
The first problem concerns the uniqueness of the analytic 
continuation from $n \in \mN$ to $n \in \mR$.
Even if all values of $\phi_N(n)$ are provided 
for $n=1,2,\ldots$, in general, 
it is not possible to determine the analytical 
continuation from $n \in \mN$ to $n \in \mR \ ({\rm or} \ \mC)$
uniquely. Carlson's theorem guarantees that 
the continuation in the right half complex plane 
is unique if the growth rate of 
$\left |\left \langle Z^n \right \rangle^{1/N} \right |$ 
is upper-bounded by $O(e^{\pi |n|})$ \cite{Titchmarsh1939}. 
However, this sufficient condition is not satisfied 
even in the case of the famous Sherrington--Kirkpatrick (SK) 
model that describes a fully connected Ising spin glass system 
\cite{SK1975}, 
for which $\left |\left \langle Z^n \right \rangle^{1/N}\right |$ 
scales as $O(e^{C|n|^2})$, where $C$ is a certain constant. 
van Hemmen and Palmer conjectured
that this might be related to the failure of the replica symmetric
(RS) solution of the SK model at low temperatures, 
although further exploration in this direction is difficult 
\cite{vanHemmen1978}. 

The second issue is the possible breaking of the analyticity of $\phi(n)$. 
Even if the uniqueness of the analytic continuation from $n \in \mN$ to 
$n \in \mR \ ({\rm or} \ \mC)$ is guaranteed for $\phi_N(n)$ of finite $N$, 
the analyticity of $\phi(n)=\lim_{N \to \infty} \phi_N(n)$ can be broken. 
This implies that if the analyticity breaking occurs at a certain 
point $n=n_c < 1$, the analytically continued expression 
based on $\phi(n)$ will lead to an incorrect solution for
$ n < n_c$. In an earlier study, the authors developed an exact 
expression of $\left \langle Z^n \right \rangle$
for a grand canonical version of the discrete random energy model 
(GCDREM) \cite{Mou1,Mou2} of finite system size \cite{OK2004}.
The expression indicates that the speculated scenario 
does occur in this model as $N \to \infty$. 
In addition, the model satisfies the sufficient condition of 
Carlson's theorem, guaranteeing the uniqueness of 
the analytical continuation 
for finite $N$. 
These points imply that the analyticity breaking with respect 
to the replica number $n$ is the origin of the one-step replica 
symmetry breaking (1RSB) observed in GCDREM. 
However, as GCDREM is a relatively unfamiliar model, 
the obtained result may be regarded as anomalistic 
and therefore might be taken as being less significant. 

The purpose of this paper is to address this 
view. More precisely, we develop 
an exact expression of $\left \langle Z^n \right \rangle$
that is applicable to generic random energy 
models (REMs) \cite{Derrida1981} of finite system size. 
The expression, which is a generalization of 
that developed for GCDREM, indicates that 
analyticity breaking with respect to the replica number 
occurs in standard REMs as well and is the origin of 1RSB. 

This paper is organized as follows. 
In the next section, we introduce the model. 
In order to guarantee the uniqueness of analytical continuation, 
we will mainly consider the canonical version of the discrete random 
energy model (CDREM). However, a similar approach 
is also applicable for standard continuous REM,
although the uniqueness of the analytical 
continuation from $n \in \mN$ to $n \in \mC$ is not guaranteed
by Carlson's theorem. 
This is shown in \ref{REM_appendix}. 
In section 3, the main part of the paper, 
an exact expression of $\left \langle Z^n \right \rangle$ is developed. 
In section 4, the developed expression is used to examine how 
the analyticity of $\left \langle Z^n \right \rangle$ can be 
broken as $N \to \infty$. 
The utility of the expression for assessing 
the moment of REMs of finite size is also shown. 
The final section is devoted to a summary.

\section{Model definition}
The canonical discrete random energy model (CDREM), 
which we will focus on here, is defined as follows. 
Suppose that $N$ and $M$ are natural numbers
where $\alpha=M/N$.
The energy for each of $2^N$ states $A=1,2,\ldots,2^N$ 
is determined as an independent
sample from an identical distribution 
\begin{eqnarray}
P(E)=2^{-M}
\left(
\begin{array}{c}
M\\
\frac{1}{2}M+E
\end{array}
\right),\label{prob}
\end{eqnarray}
where $E$ is limited to $-M/2,-M/2+1,\ldots,M/2$. 
We denote a set of sampled energy values as
$\{\epsilon_A\}$. 
For each realization of $\{\epsilon_A\}$, the partition function is 
defined as 
\begin{eqnarray}
Z &=&
\sum_{A=1}^{2^N}\exp(-\beta \epsilon_A),\label{part}
\end{eqnarray}
where $\beta =T^{-1} > 0$ denotes the inverse temperature. 
Our main objective is to develop an expression of 
$\left \langle Z^n \right \rangle$, the evaluation of which is 
computationally feasible for $\forall{N}, \forall{M}$ 
and $\forall{n} \in \mC$, 
where $\left \langle \cdots \right \rangle$, in this case,
represents the average with respect to $\{\epsilon_A\}$. 

Before proceeding further, it is worth explaining why this model has been 
selected. 
A distinctive property of this model is that 
the possible energy values are lower-bounded by $-M/2$, 
which makes it possible to upper-bound the absolute value
of a modified moment as 
\begin{eqnarray}
&&\left |\left \langle \left (e^{-\beta M/2} Z\right )^n 
\right \rangle^{1/N} \right |
\le \left \langle \left (e^{-\beta M/2} Z\right )^{{\rm Re}(n)} 
\right \rangle^{1/N}
=\left \langle \left ( 
\sum_{A=1}^{2^N} e^{-\beta(M/2+\epsilon_A)} \right )^{{\rm Re}(n)} 
\right \rangle^{1/N} \cr
&&
\le \left \langle \left ( 
\sum_{A=1}^{2^N} 1 \right )^{{\rm Re}(n)} 
\right \rangle^{1/N}
=2^{{\rm Re}(n)} < O(e^{\pi |n|}),
\label{carlson}
\end{eqnarray}
for any finite $N$. 
Consider an analytic function $\Psi(n;N)$ which satisfies
the condition $|\Psi(n;N)| < O(e^{\pi |n|})$. Carlson's theorem 
ensures that if 
$\left |\Psi(n;N)-\left \langle \left (e^{-\beta M/2} Z\right )^n 
\right \rangle^{1/N} \right |=0$
holds for $n=0,1,2\ldots$, 
$\Psi(n;N)$ is identical to 
$\left \langle \left (e^{-\beta M/2} Z\right )^n 
\right \rangle^{1/N}$ over the right half complex plane of $n$. 
The identity for $n=0$ trivially holds if $\Psi(0;N)=1$. 
Since $e^{-\beta M/2}$ is a non-vanishing constant
and $\left \langle \left (e^{-\beta M/2} Z\right )^n 
\right \rangle^{1/N} $ is a single-valued function, 
this implies that 
the analytical
continuation of $\left \langle Z^n \right \rangle^{1/N}$
from $n=1,2,\ldots$ to $n \in \mC$ 
is uniquely determined
in this model as long as $N$ is finite. 
This property is useful for examining intrinsic 
mathematical problems of RM because 
we can exclude the possibility of
multiple analytical continuations when 
certain anomalous behavior is observed for 
$\left \langle Z^n \right \rangle$. 
This is the main reason why we have selected CDREM as the 
objective system in the current paper. 

The technique developed in the next section is also 
applicable to generic REMs, 
as shown in \ref{REM_appendix}. 
Unfortunately, we cannot use Carlson's theorem 
for guaranteeing the uniqueness of the analytical continuation 
for the original continuous REM \cite{Derrida1981}, 
since $\left |\left \langle Z^n \right \rangle^{1/N}
\right |$ is not upper-bounded by $e^{\pi |n| }$. 
However, the analysis in \ref{REM_appendix} indicates that the behavior of 
the continuous REM in the thermodynamic limit 
is qualitatively the same as that of CDREM; 
in both systems, $\phi(n)$ for $n=1,2,\ldots$ is described by 
either of two replica symmetric solutions one of which is 
proportional to $n$ while the other is nonlinear with respect to $n$, 
and the replica symmetry breaking solution bifurcates from 
the nonlinear solution at a certain critical number $0<n_c < 1$
for $n < n_c$ when the temperature is sufficiently low. 
This implies that, for a wide class of random energy models, 
the origin of replica symmetry breaking 
is not the multiple possibilities of analytical continuation 
but the analyticity breaking 
with respect to the replica number
that occurs in the thermodynamic limit. 

\section{Exact expression of $\left \langle Z^n \right \rangle$ }
The expression of the partition function 
\begin{eqnarray}
Z=\sum_{A=1}^{2^N} \exp(-\beta \epsilon_A)
=\sum_{i=0}^{M} n_i \exp(-\beta  E_i)
=\omega^{-\frac{M}{2}}\sum_{i=0}^{M} n_i \omega^i \ \ 
(\omega\equiv e^{-\beta}), 
\end{eqnarray}
is the basis of our analysis, where
$n_i$ $(i=0,1,\ldots,M)$ represents the number of states 
with energy $E=i-M/2$. 
A remarkable property of DREM is that the probability 
distribution of $\{n_i\}$ has 
a feasible form
as a multinomial distribution 
\begin{eqnarray}
P(\{n_i\})=\delta_{\sum_{i=0}^M n_i,2^N}\frac{2^N! }{n_0!\cdots n_M!}
\prod_{i=0}^M \{P(E_i)\}^{n_i}, 
\label{multinomial}
\end{eqnarray}
where $\delta_{m,n}=1$ if $m=n$ and $0$ otherwise. 
This makes it possible to assess the moments 
$\left \langle Z^n \right \rangle$ directly from 
$\{n_i \}$ without referring to the full energy configuration 
$\{\epsilon_A\}$, yielding
\begin{eqnarray}
\left\langle Z^n\right\rangle
&=& \sum_{n_0=0}^{\infty}
\sum_{n_1=0}^{\infty}
\cdots
\sum_{n_M=0}^{\infty} P(\{n_i\}) Z^n \cr 
&=&
\sum_{n_0=0}^{\infty}
\sum_{n_1=0}^{\infty}
\cdots
\sum_{n_M=0}^{\infty}
\{P(E_0)\}^{n_0}
\{P(E_1)\}^{n_1}
\cdots
\{P(E_M)\}^{n_2}\ \nonumber\\
&&\times \delta_{\sum_{i=0} n_i,2^N}\frac{2^N!}{n_0!\cdots n_M!}\ 
Z^n.\label{CDREM}
\end{eqnarray}

\begin{figure}
       \centerline{\includegraphics[width=8cm]{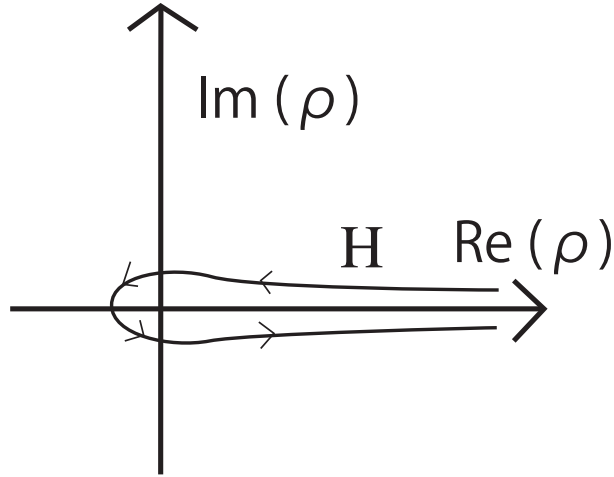}}
\caption{Integration contour.}
\label{cont}
\end{figure}

Two identities are useful for evaluating eq. (\ref{CDREM}):
\begin{eqnarray}
c^n
=
\frac{\int_H (-\rho)^{-n-1}e^{-c\rho}d\rho}{\tilde \Gamma (-n)}, \ \ (c>0,\tilde\Gamma (n)\equiv -2 \img \sin n\pi \Gamma (n)),\label{cn}
\end{eqnarray}
and 
\begin{eqnarray}
\delta_{m,n}
=
\frac{1}{2 \pi \img }\oint z^{m-n-1}dz,
\label{delta}
\end{eqnarray}
where $\img =\sqrt{-1}$. 
In eq. (\ref{cn}), 
$\Gamma(z)=\img/(2 \sin z \pi ) \int_H (-\rho)^{z-1}e^{-\rho} d\rho$ 
represents
the Gamma function and the integration contour $H$ is provided as 
shown in figure \ref{cont}. 
The integration contour of eq. (\ref{delta}) 
is a single closed loop surrounding the origin. 
Inserting eqs. (\ref{cn}) and (\ref{delta}) into 
eq. (\ref{CDREM}) yields 
\begin{eqnarray}
\left\langle Z^n\right\rangle 
&=&
\sum_{n_0=0}^{\infty}
\sum_{n_1=0}^{\infty}
\cdots
\sum_{n_M=0}^{\infty}
\{P(E_0)\}^{n_0}
\{P(E_1)\}^{n_1}
\cdots
\{P(E_M)\}^{n_2}\ 
\frac{2^N!}{n_0!\cdots n_M!}\ \nonumber\\
&&\frac{
\omega^{-\frac{nM}{2}}
}{
\tilde\Gamma (-n)
}
\int_H d\rho(-\rho)^{-n-1}e^{-(\sum_{i=0}^{M} n_i \omega^i)\rho}
\frac{1}{2\pi \img }\oint dz z^{\sum_{i=0}^{M}n_i -2^N-1} \cr
&=&
\frac{
\omega^{-\frac{nM}{2}}
}{
\tilde\Gamma (-n)
}
\int_H d\rho(-\rho)^{-n-1}
\frac{(2^N)!}{2\pi \img }\oint dz z^{-2^N-1}
\left(
\sum_{n_0=0}^{\infty}\frac{1}{n_0!}\{zP(E_0)e^{-\rho}\}^{n_0}
\right)\nonumber\\
&&\left(
\sum_{n_1=0}^{\infty}\frac{1}{n_1!}\{zP(E_1)e^{-\omega\rho}\}^{n_1}
\right)
\cdots\left(
\sum_{n_M=0}^{\infty}\frac{1}{n_M!}\{zP(E_M)e^{-\omega^M\rho}\}^{n_M}
\right)\nonumber\\
&=&
\frac{
\omega^{-\frac{nM}{2}}
}{
\tilde\Gamma (-n)
}
\int_H d\rho(-\rho)^{-n-1}
\frac{(2^N)!}{2\pi \img }\oint dz z^{-2^N-1}
e^{z\sum_{i=0}^M P(E_i)e^{-\omega^i\rho}}.
\end{eqnarray}
Finally, applying the residue theorem to the integration 
with respect to $z$, we obtain 
\begin{eqnarray}
\left\langle Z^n\right\rangle 
&=&
\frac{
\omega^{-\frac{nM}{2}}
}{
\tilde\Gamma (-n)
}
\int_H d\rho(-\rho)^{-n-1}
\left(\sum_{i=0}^M P(E_i)e^{-\omega^i\rho}\right)^{2^N}, \label{exactmoment}
\end{eqnarray}
which exactly holds for any complex value of $n$. 
This is the main result of this paper.

Several points are noteworthy here. 
The first issue is a relation to an earlier study. 
In ref. \cite{GardnerDerrida1989}, 
Gardner and Derrida provided an expression somewhat 
similar to eq. (\ref{exactmoment}) for non-integer moments 
of the partition function of the continuous REM. 
Actually, the present expression of eq. (\ref{exactmoment}) 
can be regarded as a generalization of eq. (7) of ref. 
\cite{GardnerDerrida1989}, for which, however, 
tractability for finite $N$ is not emphasized.  
The link of the two expressions is shown in \ref{REM_appendix}. 
The second is the computational cost 
for evaluating eq. (\ref{exactmoment}). 
For each $\rho \in \mC$, the integrand of 
eq. (\ref{exactmoment}) can be evaluated 
with $O(N)$ computations provided 
$\alpha = M/N \sim O(1)$. 
On the other hand, the length of the integration contour $H$ is infinite, 
which may seem an obstacle for the numerical evaluation. 
However, for relatively large $N$, the integrand rapidly vanishes 
on the contour $H$ as $|\rho| \to \infty$,
guaranteeing that numerical deviation is practically negligible 
even if we approximate $H$ by a path of a finite length. 
Therefore, eq. (\ref{exactmoment}) can be evaluated with 
a feasible computational cost. 
The third point is the similarity to 
an approach in information theory 
sometimes termed the {\em method of types} \cite{CsiszarKorner1981}. 
In this method, the performance of various codes consisting of exponentially many codewords
is evaluated by classifying possible events when the codes are 
randomly generated by varieties of empirical distributions, 
termed {\em types}.  
The key to this method is accounting for 
the fact that the number of types grows only polynomially with respect to 
the code length while the number of the codewords increases exponentially. 
As a consequence, 
the relative weight of the types 
concentrates at a certain typical value at an exponentially fast rate, 
which considerably simplifies the performance analysis. 
In the current case, the set of occupation numbers $\{n_i\}$ 
is analogous to the types in information theory. 
However, in spite of this analogy, objective quantities are 
usually evaluated by upper- and lower-bounding schemes
in information theory. Therefore, the technique
developed in the present paper may 
serve as a novel scheme for analyzing various codes in 
information theory. The final issue relates to the relationship to a model
examined in an earlier study. In ref. \cite{OK2004}, the authors
offered a formula similar to eq. (\ref{exactmoment})
for a grand canonical version of the discrete random energy model 
(GCDREM). Unlike CDREM, GCDREM is defined by {\em independently }
generating the occupation number $n_i$ of each energy level
from the beginning. 
In the present framework, this can be characterized by a factorizable 
distribution of $\{n_i\}$, 
\begin{eqnarray}
P(\{n_i\})= e^{-2^{N}} \prod_{i=0}^M 
\frac{\left (-P(E_i)2^{N} \right)^{n_i}}{n_i!}, 
\label{GCDREM}
\end{eqnarray}
which offers a feasible expression 
\begin{eqnarray}
  \label{consistent}
  \left\langle Z^n \right\rangle
  =
\frac{
\omega^{-\frac{nM}{2}}
}{
\tilde\Gamma (-n)
}
\int_H d\rho (-\rho)^{-n-1}\exp{[-\sum_{i=0}^{M}
(1-e^{-\omega^i \rho})P(E_i)2^N]}. 
\end{eqnarray}
This is the counterpart of eq. (\ref{exactmoment}). 
Since the distributions of $n_i$ are independent, 
the derivation of eq. (\ref{consistent}) is easier 
than that of eq. (\ref{exactmoment}). 
In GCDREM, the total number of states ${\cal N}=\sum_{i=0}^M n_i$
fluctuates from sample to sample 
as $n_i$ is independently generated 
at each energy level.
The ratio between its standard deviation and expectation vanishes
as $2^{-N/2}$ and therefore the effect of the 
statistical fluctuation is practically negligible 
when $N$ is reasonably large. 
Nevertheless, eq. (\ref{GCDREM}) implies that 
an ill-posed sample $n_i=0$ $(i=0,1,\ldots,M)$ can 
be generated with a probability $2^{-N}$, 
which is not adequate for representing REMs of small system size. 
Therefore, the development of eq. (\ref{exactmoment}) 
beyond eq. (\ref{consistent}) is important 
for REMs of finite size.

\section{Applications }
In this section, we demonstrate the utility of the developed expression (\ref{exactmoment}) 
by considering two applications. 
\subsection{Thermodynamic limit and breaking of analyticity}
RM indicates that CDREM exhibits the following behavior
in the thermodynamic limit $N,M \to \infty$ keeping 
$\alpha=M/N $ finite. For details, see \ref{replica_appendix}. 

\begin{figure}
       \centerline{\includegraphics[width=12cm]{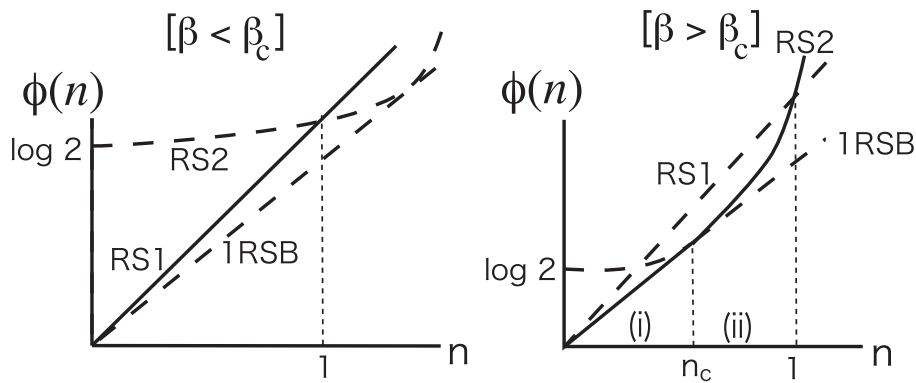}}
\caption{Behavior of $\phi(n)$ predicted by RM. }
\label{phi_CDREM}
\end{figure}

Under the RS ansatz, RM yields two solutions
\begin{eqnarray}
\phi_{\rm RS1}(n)=n \left (\log 2+\alpha \log \left (\cosh \left (
\frac{\beta}{2} \right ) \right ) \right ), 
\label{RS1}
\end{eqnarray}
and 
\begin{eqnarray}
\phi_{\rm RS2}(n)=\log 2+\alpha \log \left (\cosh \left (
\frac{n \beta}{2} \right ) \right ). 
\label{RS2}
\end{eqnarray}
For $\forall{\alpha}$ and $\forall{\beta}$, 
these solutions agree at $n=1$. 
Let us denote a critical inverse temperature as
\begin{eqnarray}
\beta_c= \left \{
\begin{array}{ll}
\infty, &  \alpha \le 1\cr
\log \left (1-h_2^{-1}(1-\alpha^{-1}) \right )
-\log \left (h_2^{-1}(1-\alpha^{-1}) \right ), 
& \alpha > 1
\end{array}
\right . , 
\label{criticaltemp}
\end{eqnarray}
where $h_2^{-1}(y)$ is the inverse function of the 
binary entropy $h_2(x)=-x \log_2 (1-x)
- (1-x)\log_2 (1-x)$ for $0<x<1/2$. 
For $\beta < \beta_c$, 
$\phi(n)$ can be described by $\phi_{\rm RS1}(n)$ and $\phi_{\rm RS2}(n)$. 
More precisely, there exists $\exists{n_{\rm RS}} > 1$
such that $\phi(n)=\phi_{\rm RS2}(n)$ for $n > n_{\rm RS}$ and 
$\phi(n)=\phi_{\rm RS1}(n)$ for $n < n_{\rm RS}$. 
On the other hand, for $\beta > \beta_c$, $\phi(n)$ cannot be entirely covered 
by the RS solutions. In this ``low temperature''
phase, $\phi(n)=\phi_{\rm RS2}(n)$ holds for $n > n_c=\beta_c/\beta$. 
However, this solution becomes inadequate for $n < n_c$ 
since the convexity condition $\left (\partial /\partial n \right )
\left (n^{-1} \phi_{\rm RS2}(n) \right ) \ge 0$ is not satisfied. 
Therefore, we have to construct a novel solution for $n < n_c$ 
taking the breaking of replica symmetry into account, 
which provides the 1RSB solution
\begin{eqnarray}
\phi_{\rm 1RSB}(n)=\frac{n\alpha \beta}{2} \tanh \frac{\beta_c}{2}. 
\label{1RSB}
\end{eqnarray}
The behavior of $\phi(n)$ is shown in figure \ref{phi_CDREM}. 

As long as $N$ is finite, 
$\phi_N(n)$ is generally analytic with respect to $n$ 
and Carlson's theorem guarantees the 
uniqueness of the analytical continuation 
from $n \in \mN$ to $n \in \mC$ for $\phi_N(n)$ 
of CDREM. 
This implies that the transitions between multiple solutions mentioned above
are highly likely to be due to the breaking of analyticity 
that possibly occurs in the limit $\phi(n)=\lim_{N \to \infty}
\phi_N(n)$. The expression of eq. (\ref{exactmoment})
can be used to examine the origin of such 
potential analyticity breaking.

We here focus on the breaking for 
the transition between $\phi_{\rm RS2}(n)$ and $\phi_{\rm 1RSB}(n)$, 
which can be regarded as the origin of 1RSB.
For this purpose, we assume $\alpha >1$ and $\beta > \beta_c$, and
consider only the region $n < 1$ in which $n_c=\beta_c/\beta < 1$ 
is included. Under such conditions, integration by parts 
transforms eq. (\ref{exactmoment}) to 
\begin{eqnarray}
\left\langle Z^n\right\rangle 
&=&
\frac{
2^N\omega^{-\frac{nM}{2}}
}{
n\tilde\Gamma (-n)
}
\int_H d\rho(-\rho)^{-n}
g(\rho)f(\rho),
\label{integratbypart}
\end{eqnarray}
where
\begin{eqnarray}
f(\rho)
=
\left(\sum_{i=0}^M P(E_i)e^{-\omega^i\rho}\right)^{2^N-1}
\label{f}
\end{eqnarray}
and
\begin{eqnarray}
g(\rho)
=
\sum_{i=0}^M P(E_i)\omega^ie^{-\omega^i\rho}.
\label{g_func}
\end{eqnarray}
Next, we deform the contour $H$ to a set of straight lines 
along the real half axis as $\widetilde{H}: \infty + 
\img \epsilon \to + \img \epsilon 
\to -\img \epsilon \to \infty -\img \epsilon$, where 
$\epsilon$ is an infinitesimal number. 
Since the contribution around the origin vanishes for ${\rm Re}(n)< 1$, 
only the contribution due to the difference between the branches of $\rho^{-n}$ 
remains as
\begin{eqnarray}
\left\langle Z^n\right\rangle 
&=&
\frac{
2^N\omega^{-\frac{nM}{2}}
}{
\Gamma (1-n)
}
\int_0^{\infty} d\rho \rho^{-n}
g(\rho)f(\rho).
\label{integrationonreal}
\end{eqnarray}
Note that the function $\tilde\Gamma(z)$ is replaced with 
the ordinary gamma function $\Gamma(z)$ in this expression 
due to the difference between the two branches.
We now evaluate eq. (\ref{integrationonreal}),
converting the relevant variables as 
\begin{eqnarray}
x=\frac{i}{N\alpha},\ \ y=\frac{1}{N\alpha}\log \rho,
\end{eqnarray}
which is useful for taking the thermodynamic limit $N,M \to \infty$
keeping $\alpha=M/N$ finite. 
This yields an expression of the moment 
\begin{eqnarray}
\left\langle Z^n\right\rangle 
&=&
\frac{
2^N\omega^{-\frac{nN \alpha}{2}}N\alpha\beta
}{
\Gamma (1-n)
}
\int_{-\infty}^{\infty} dy e^{(1-n)N\alpha\beta y}
g(e^{N\alpha\beta y})f(e^{N\alpha\beta y}).
\label{asymptotic_exactmoment}
\end{eqnarray}
The analysis shown below indicates that 
$\alpha(x-1/2)$ and $\alpha(y-1/2)$ can be
interpreted as possible values of energy density $N^{-1} E_i$ and 
free energy density $-(N \beta )^{-1} \log Z$, respectively. 

An asymptotic form of the double exponential function for $N \gg 1$
\begin{eqnarray}
e^{-e^{-Nu}}
\sim \left (1-e^{-Nu } \right ) \Theta(u),
\label{doubleexp}
\end{eqnarray}
where $\Theta(u)=1$ for $u>0$ and $0$ otherwise, 
plays a key role for assessing eq. (\ref{asymptotic_exactmoment})
in the thermodynamic limit. 
We first evaluate the asymptotic expression of $f(e^{N\alpha\beta y})$ 
using this formula. Eq. (\ref{doubleexp}), in conjunction with 
an assessment by the saddle point method, yields
\begin{eqnarray}
&& \sum_{i=0}^M P(E_i)e^{-\omega^i\rho}\nonumber\sim
M \int_0^1 dx \ e^{N\alpha \left (h(x)-\log{2} \right )}\  e^{-e^{-N\alpha\beta(x-y)}}\nonumber\\
&&\sim 
M \int_0^1 dx \ e^{N\alpha \left (h(x)-\log{2} \right )}\  
(1-e^{-N\alpha\beta(x-y)})\Theta \left (x-y \right ) \cr
&& \sim 
\left \{
\begin{array}{ll}
1-e^{N\alpha \left (\log(1+\omega)+\beta y-\log{2} \right )}, & y<x_c, \cr
1-e^{N\alpha \left (h(x)-\log{2} \right )}, & x_c <y< 1/2 , \cr
0, & 1/2 < y , 
\label{fasympt}
\end{array}
\right . 
\end{eqnarray}
where $x_c=e^{-\beta}/(1+e^{-\beta})< 1/2$  and 
$h(x)\equiv -x\log{x}-(1-x)\log{(1-x)}$.
Therefore, $f(e^{N\alpha\beta y})$ can be evaluated as
\begin{eqnarray}
f(e^{N\alpha\beta y})
&=& e^{(2^N-1)
\log{\left (\sum_{i=0}^M P(E_i)e^{-\omega^ie^{N\alpha\beta y}}\right )}} 
\sim 
e^{-e^{N {\cal F}(y)}}\Theta \left (\frac{1}{2}-y \right ), 
\label{fasymp2}
\end{eqnarray}
where
\begin{eqnarray}
{\cal F}(y) =
\left\{
\begin{array}{ll}
(1-\alpha)\log{2}+\alpha {\widetilde h}(\beta) + \alpha \beta y, &
 y<x_c,  \cr 
(1-\alpha)\log{2}+\alpha h(y), & x_c< y < 1.
\end{array}
\right.
\end{eqnarray}
Here, ${\widetilde h}(\beta)$ represents the Legendre transformation 
of the function ${h}(x)$ as 
${\widetilde h}(\beta)
=
\log\left (2\cosh \left (\beta/2 \right )   \right )
-\beta/2
=
\log{(1+\omega)}$.
These indicate that in the limit $N\rightarrow \infty$
eq. (\ref{fasymp2}) can be reduced to the expression 
\begin{eqnarray}
f(e^{N\alpha\beta y})
&\sim &
\left (1-e^{N{\cal F}(y)} \right )\Theta (x^{*}-y)
\sim \Theta (x^{*}-y),
\end{eqnarray}
where $x^{*}=h_2^{-1} \left (1-\alpha^{-1} \right )
=e^{-\beta_c}/(1+e^{-\beta_c})$. 
Similarly, $g(e^{N\alpha\beta y})$ can be evaluated as
\begin{eqnarray}
g(e^{N\alpha\beta y})
\sim 2^{-N}e^{N \left (-\alpha \beta y +{\cal F}(y)\right )} \Theta(1-y).
\label{gfunc}
\end{eqnarray}
From these, the moment can be asymptotically expressed as
\begin{eqnarray}
\left\langle Z^n\right\rangle 
&\sim &
\frac{
2^N\omega^{-\frac{nM}{2}}N\alpha\beta
}{
\Gamma (1-n)
}
\int_{-\infty}^{x^{*}} dy\ e^{(1-n)N\alpha\beta y}\ 
g(e^{N\alpha\beta y}) \cr
&\sim &
\frac{
\omega^{-\frac{nM}{2}}N\alpha\beta
}{
\Gamma (1-n)
}
\int_{-\infty}^{x^{*}} dy\  e^{N {\cal G}(y)}
\label{calG}
\end{eqnarray}
where
\begin{eqnarray}
{\cal G}(y)
=-n\alpha \beta y +{\cal F}(y).
\label{func_calG}
\end{eqnarray}

Based on eq. (\ref{calG}), the thermodynamic limit 
$\phi(n)=\lim_{N \to \infty}N^{-1}\log \left\langle Z^n\right\rangle $
can be assessed by examining the profile of ${\cal G}(y)$. 
For the case of $\beta > \beta_c$ and $\alpha >1$, 
which we currently focus on, 
$x^* =e^{-\beta_c}/(1+e^{-\beta_c}) > x_c
=e^{-\beta}/(1+e^{-\beta})$ is guaranteed.
This implies that the asymptotic behavior 
can be classified into two cases depending on whether or not
$y_c=e^{-n\beta}/(1+e^{-n\beta})$, which maximizes ${\cal G}(y)$, is 
included in the integration range $y < x^{*}$: 
\begin{figure}
       \centerline{\includegraphics[width=10cm]
                                   {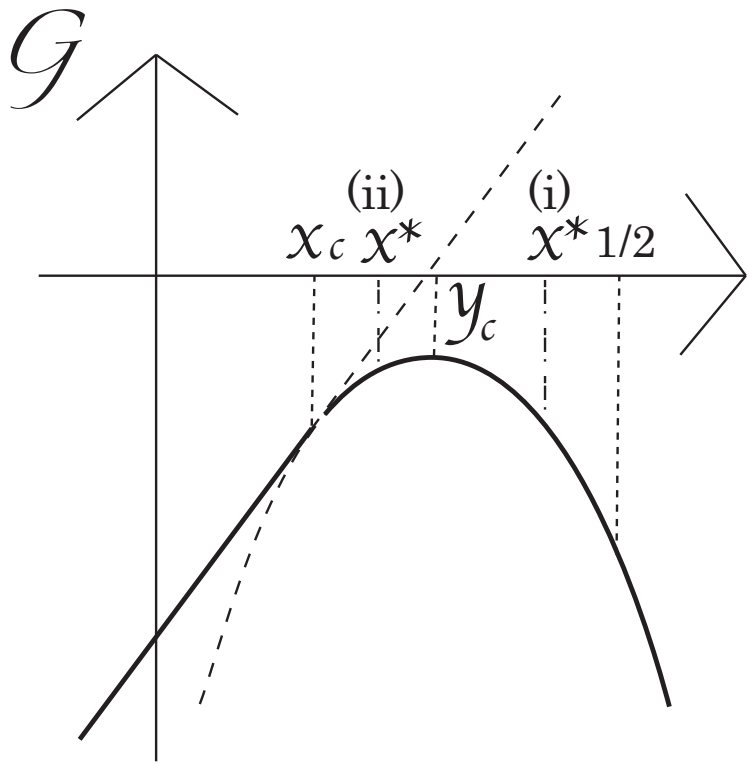}}
\caption{Schematic figure of ${\cal G}(y)$
for $\beta>\beta_c$ and $\alpha > 1$.  
The maximum value in the integration range 
$y < x^*$ determines the asymptotic behavior 
$\phi(n)=\lim_{N \to \infty} N^{-1}\log 
\left \langle Z^n \right \rangle $.}
\label{g}
\end{figure}

\begin{enumerate}
\renewcommand{\labelenumii}{(\roman{enumii})} 
\item  $y_c < x^{*}$\\
As $y_c$ and $x^{*}$ are parameterized
as $y_c=e^{-n\beta}/(1+e^{-n \beta})$ and 
$x^{*}=e^{-\beta_c}/(1+e^{-\beta_c})$, respectively, 
this holds for $n > n_c \equiv \beta_c/\beta$. 
In this case, ${\cal G}(y)$ is maximized at $y=y_c$, as shown in figure 
\ref{g},
which means eq. (\ref{calG}) can be assessed as
$\left\langle Z^n\right\rangle  \sim \exp \left [N 
\left ({\cal G}(y_c) + n \alpha \beta /2 \right )\right ]$. 
This yields the thermodynamic limit 
\begin{eqnarray}
\phi(n)=\lim_{N\rightarrow\infty}
\frac{1}{N}\log{\left\langle Z^n\right\rangle }
=\log{2}+\alpha \log\left (\cosh\left (\frac{n\beta}{2} \right ) \right ), 
\end{eqnarray}
which coincides with $\phi_{\rm RS2}(n)$. 
\item $y_c > x^{*}$\\
This holds for $n < n_c$. 
Figure \ref{g} indicates that ${\cal G}(y)$ is maximized at 
$y=x^{*}$, which is the right-end point of the integration range. 
In this case, the asymptotic behavior is given as 
\begin{eqnarray}
\phi(n)=\lim_{N\rightarrow\infty}
\frac{1}{N}\log{\left\langle Z^n\right\rangle }=
n\alpha\beta\left (\frac{1}{2}-x^{*} \right )
=
\frac{n\alpha\beta}{2}\tanh{\frac{\beta_c}{2}}.
\end{eqnarray}
This is identical to $\phi_{\rm 1RSB}(n)$. 
\end{enumerate}

The analyticity breaking at $n=n_c$ for $\phi(n)=\lim_{N \to \infty}
\phi_N(n)=\lim_{N \to \infty} N^{-1}\log 
\left \langle Z^n \right \rangle$
can be interpreted as follows \cite{BouchaudMezard1997,OK2004}. 
In the low temperature phase of $\beta > \beta_c$, 
it is considered that REM of each sample 
is dominated by a few states of minimum energy. 
This means that the free energy density in this phase can be 
roughly expressed as $-(N\beta)^{-1} \log Z \simeq N^{-1}\epsilon_{\rm min}$, 
where $\epsilon_{\rm min}$ denotes the minimum value of 
the $2^N$ energy states $\epsilon_1, \epsilon_2, \ldots, \epsilon_{2^N}$. 
The theory of extreme value statistics \cite{Gumbel1958} 
indicates that 
$y=M^{-1} \epsilon_{\rm min} +1/2$,  
obeys a Gumbel-type distribution, $P(y)$, 
which in the current system is characterized as
\begin{eqnarray}
\frac{1}{N} \log P(y) =
\left \{ 
\begin{array}{ll}
-\infty, & y > x^*, \cr
(1-\alpha)\log{2}+\alpha h(y), & 0 < y < x^*,  
\end{array}
\right .
\label{Gumbel}
\end{eqnarray}
as $N$ tends to infinity, where $x^*=h_2^{-1}(1-\alpha^{-1})$ represents 
the typical value of $y$. The physical implication of this 
behavior is that 
it is very rare for $y$ to fluctuate in the right direction because
$y$ is given by the minimum of exponentially 
many energy values, 
while the fluctuation in the left direction obeys normal-type
large deviation statistics described by a finite rate function. 
Eq. (\ref{Gumbel}) indicates that 
if $y_c < x^{*}$ holds, which is the case for $n > n_c$, 
then $y=y_c=e^{-n\beta}/(1+e^{-n \beta})$, 
which corresponds to a rare low minimum energy, 
dominates $\left \langle Z^n \right \rangle 
=N\alpha \beta \omega^{-\frac{nM}{2}}
\int dy e^{-N n \alpha \beta y} P(y)$, 
yielding $\phi(n)=\phi_{\rm RS2}(n)$. 
On the other hand, for $n < n_c$, $\left \langle Z^n \right \rangle $
is dominated by $y=x^*$, which corresponds to the typical value 
of the minimum energy, and the thermodynamic limit is provided as
$\phi(n)=\phi_{\rm 1RSB}(n)$. These indicate that the origin 
of the 1RSB solution is a singularity of the distribution 
of the minimum energy, which arises in the thermodynamic limit 
and is characterized as eq. (\ref{Gumbel}). 
The formula (\ref{exactmoment}) makes it possible to
more precisely describe the analyticity breaking of $\phi(n)$ brought about by 
this singularity.

\subsection{Numerical assessment of the moment}
The utility of eq. (\ref{exactmoment}) is not limited to analysis in
the thermodynamic limit; this formula is also useful for 
the numerical assessment of the moment for systems of finite size. 

A representative approach to evaluate $\left \langle Z^n \right \rangle $
is to  numerically average $Z^n$ by sampling $n_0,n_1,\ldots, n_M$
from eq. (\ref{multinomial}). 
This can be efficiently performed as follows 
(also shown in Appendix C of ref. \cite{OK2004}). 

To generate $n_0,n_1,\ldots, n_M$ following eq. (\ref{multinomial}), 
we first sample $n_0$ from a binomial distribution 
\begin{eqnarray}
P(n_0)=
\frac{2^N! }{\left (2^N -n_0 \right )!n_0!}
p_0^{n_0}\left (1-p_0 \right )^{2^N-n_0}, 
\label{p0}
\end{eqnarray}
where $p_0=P(E_0)$. For a generated $n_0$, we next sample $n_1$
from a conditional binomial distribution 
\begin{eqnarray}
P(n_1|n_0)=
\frac{\left (2^N -n_0 \right )!}{\left (2^N -n_0 -n_1\right )!n_1!}
p_1^{n_1}\left (1-p_1 \right )^{2^N-n_0-n_1}, 
\label{p1}
\end{eqnarray}
where $p_1=P(E_1)/(1-P(E_0))$. We repeat this process up to 
$n_{M-1}$; namely, for a generated set of $n_0,n_1,\ldots,n_{i-1}$, 
$n_i$ is sampled from a conditional binomial distribution
\begin{eqnarray}
P(n_i|n_0,n_1,\ldots,n_{i-1})
=\frac{\left (2^N -\sum_{k=0}^{i-1}n_k \right )!}
{\left (2^N -\sum_{k=0}^{i}n_k \right )! n_i! }
p_i^{n_i}\left (1-p_i \right )^{2^N-\sum_{k=0}^{i-1}n_k}, 
\label{pi}
\end{eqnarray}
where $p_i=P(E_i)/(1-\sum_{k=0}^{i-1}P(E_k))$. 
After generating $n_0,n_1,\ldots,n_{M-1}$ in this manner, 
$n_M$ is given as $n_M=2^N-\sum_{i=0}^{M-1}n_i$, 
which guarantees satisfaction of the strict constraint $\sum_{i=0}^M n_i=2^N$. 

Since each step can be performed by an $O(1)$ computation, 
the necessary computational cost for this generation is $O(N)$ per sample, 
which is feasible. 
However, a problem remains in the evaluation of 
$\left \langle Z^n \right \rangle$ by this 
approach based on naive sampling. 

The analysis shown in the preceding subsection 
indicates that in the low temperature phase 
$\beta > \beta_c$, $\left \langle Z^n \right \rangle$ 
of $n > n_c$ is dominated by samples of a rare low minimum energy value
in the limit $N \to \infty$.  
This implies that accurate assessment of such moments
requires an exponentially large number of samples since
the probability of generating the dominant samples decreases
exponentially with respect to $N$. On the other hand, 
the analysis also means that eq. (\ref{exactmoment})
can be accurately evaluated by numerical integration.  
This is because the integrand of eq. (\ref{integrationonreal}) is 
practically localized in a finite interval due to 
the properties of large deviation statistics 
and therefore truncating the integration range to a finite interval 
yields little numerical error.

\begin{figure}
       \centerline{\includegraphics[width=8cm]
                                   {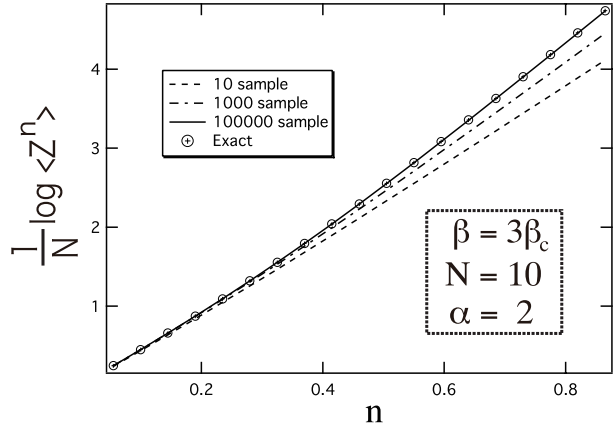}}
\caption{$\left \langle Z^n \right \rangle$ 
assessed by numerical integration of eq.(\ref{exactmoment}) 
and a naive sampling scheme for $N=10$. 
The relevant parameters are $\alpha=2$ and $\beta=3 \beta_c$, 
which corresponds to the low temperature phase. 
}
\label{sample}
\end{figure}

Figure \ref{sample} shows a comparison of the numerical evaluation of
$\left \langle Z^n \right \rangle$ between the sampling and numerical 
integration approaches for the low temperature phase of 
$\alpha=2, \beta=3 \beta_c$. In both approaches 
the system size was set to $N=10$ 
while the number of samples was varied as ${\cal N}_{\rm samples}=10, 
10^3$ and $10^6$. The figure shows that as $n$ increases, 
more samples are necessary to accurately evaluate 
$\left \langle Z^n \right \rangle$. This is consistent with the 
analysis in the preceding 
subsection indicating that $\left \langle Z^n \right \rangle$
for $n \gtsim n_c$ is dominated by samples of a rare low minimum energy. 
In the sampling approach, ${\cal N}_{\rm samples}=10^6$ 
achieves an accuracy similar to
that of the numerical integration. However, more samples are 
necessary as $N$ increases 
because the probability of generating samples 
that dominate $\left \langle Z^n \right \rangle$ decreases 
exponentially with $N$. 
On the other hand, the necessary computational cost for 
evaluating $\left \langle Z^n \right \rangle$ is not greatly 
dependent on $N$ in the numerical integration, 
demonstrating a point of superiority of eq. (\ref{exactmoment}). 

Notice that eq. (\ref{exactmoment}) holds not only for 
$n \in \mR$ but also for $n \in \mC$. 
This can be used for characterizing the breaking of 
the analyticity of $\left \langle Z^n \right \rangle$ by 
the convergence of the zeros of $\left \langle Z^n \right \rangle$ 
to the real axis on the complex $n$ plane as $N$ tends to infinity, 
which is analogous to a description of phase transitions by 
Lee and Yang \cite{LeeYang} and Fisher \cite{Fisher}. 
As far as the authors know, 
the complex zeros with respect to $n$ have little been 
examined except for a few examples \cite{OK2005,Obuchi2009} 
although zeros of the complex temperature 
or external field for typical samples of REMs were 
studied in preceding literature \cite{Mou1,Derrida1991}. 
A detailed analysis along this direction will be reported in a 
subsequent paper \cite{OK2008b}. 

\section{Summary}\label{summary}
In summary, we have developed an exact expression for the moment of 
the partition function of CDREM. 
The expression is valid not only in the thermodynamic limit 
but also for systems of finite size. 
This is useful for investigating the breaking of analyticity with 
respect to the replica number, which plays the role of 
the power of the moment, in the thermodynamic limit. 
Our approach demonstrates that the analyticity breaking 
due to a singularity in the distribution of the minimum 
energy is the origin of 1RSB observed in the low temperature phase
of CDREM. 
We have also shown that the developed expression is useful 
for numerically evaluating the moment of finite size systems. 

Although the uniqueness of analytical continuation 
from $n \in \mN$ to $n \in \mC$ is 
not guaranteed by Carlson's theorem for generic REMs, 
the behavior predicted by our approach is qualitatively 
the same as that of CDREM. This implies that
1RSB observed in a wide class of REMs 
originates from the analyticity breaking with respect 
to the replica number $n$, 
which takes place in the thermodynamic limit $N \to \infty$
after analytical continuation from $n \in \mN$ to 
$n \in \mC$ is performed for finite $N$. 

\ack
The authors appreciate anonymous referees
for their useful comments about a similarity between 
the present study and that of ref. \cite{GardnerDerrida1989}.
This work was supported by a Grant-in-Aid 
for Young Scientists (B) by JSPS (KO) 
and that on the Priority Area 
``Deepening and Expansion of Statistical Mechanical
Informatics'' by  the Ministry of Education, Culture, Sports and Science (YK). 

\appendix
\section{Exact expression of $\left \langle Z^n \right \rangle$ 
for continuous random energy model}
\label{REM_appendix}
We here provide an exact expression of $\left \langle Z^n \right \rangle$ 
for the continuous random energy model, which was 
originally introduced by Derrida in \cite{Derrida1981}.  
In this model, $2^N$ states are distributed 
in energy $E$ with the following probability, 
\begin{eqnarray}
  \label{eq:prob}
  P(E)
  =
  \frac{1}{\sqrt{\pi N J^2}}
  \exp{\left(
      -\frac{E^2}{N J^2}
    \right)}.
\end{eqnarray}
The moment of the partition function
$Z(\{E_i\})  =  \sum_{i=1}^{2^N}  \exp{\left(- \beta E_i\right)}$
is expressed as
\begin{eqnarray}
  \label{eq:conzn}
  \left \langle Z^n \right \rangle
  =
  \int_{-\infty}^{\infty}
  \left(
  \prod_{i=1}^{2^N}
  dE_i
    P(E_i)
  \right)
  Z^n\left (\{E_i\} \right ).
\end{eqnarray}
Applying eq. (\ref{cn}) to $Z^n\left (\{E_i\} \right )$ yields
\begin{eqnarray}
  \left \langle Z^n \right \rangle
  &=&
  \frac{
    1
  }{
    \tilde\Gamma (-n)
  }
  \int_H d\rho(-\rho)^{-n-1}
  \prod_{i=1}^{2^N}
  \left(
    \int_{-\infty}^{\infty}
    dE_i
    P(E_i)
    e^{-\rho\exp{\left(-\beta {E_i}\right)}}
  \right)\cr
  &=&
    \frac{
    1
  }{
    \tilde\Gamma (-n)
  }
  \int_H d\rho(-\rho)^{-n-1}
  \left(
    \int_{-\infty}^{\infty}
    dE
    P(E)
    e^{-\rho\exp{\left(-\beta {E}\right)}}
  \right)^{2^N}\cr
    &=&
    \frac{
    1
  }{
    \tilde\Gamma (-n)
  }
  \int_H d\rho(-\rho)^{-n-1}
  \left(
    \int_{-\infty}^{\infty}
    \frac{dE}{\sqrt{\pi N J^2}}
    e^{-\frac{E^2}{N J^2}-\rho\exp{\left(-\beta {E}\right)}}
  \right)^{2^N}.
\end{eqnarray}
Numerically evaluating this is computationally feasible 
for $\forall{N}$. 

Deforming the contour $H$ to $\widetilde{H}: \infty + 
\img \epsilon \to + \img \epsilon 
\to -\img \epsilon \to \infty -\img \epsilon$ with 
an infinitesimal number $\epsilon$
(putting $\rho = r e^{i(\theta-\pi)},\ \theta=-\pi, \pi$), 
in conjunction with $p > {\rm Re}(n)$ times employment of the 
integration by parts for removing the singularity at $\rho=0$,  
yields
\begin{eqnarray}
\left \langle Z^n \right \rangle
&=&\frac{1}{\Gamma(p-n)}
\int_0^\infty dr \ r^{p-n-1} \cr
&\phantom{=}&
\phantom{\frac{1}{\Gamma(p-n)}}
\times \left (-\frac{\partial}{\partial r} \right )^p
  \left(
    \int_{-\infty}^{\infty}
    \frac{dE}{\sqrt{\pi N J^2}}
    e^{-\frac{E^2}{N J^2}-r\exp{\left(-\beta {E}\right)}}
  \right)^{2^N}, 
\end{eqnarray}
which is equivalent to eq. (7) of ref. \cite{GardnerDerrida1989}. 

From eq. (\ref{eq:conzn}), the behavior of the moment 
in the thermodynamic limit can be investigated 
in a manner similar to the analysis of CDREM. 
This investigation indicates that 
the moment behaves as 
\begin{eqnarray}
  \label{eq:conabove}
\phi(n)=  \lim_{N\rightarrow\infty}
  \frac{1}{N}\log{\left\langle Z^n\right\rangle }
  =
\left \{
\begin{array}{ll}
\log{2}+n^2 \beta^2 J^2/4, & n > n_{\rm RS}, \cr
n\left(\log{2}+{\beta^2 J^2}/{4}\right), & n < n_{\rm RS}, 
\label{highT}
\end{array}
\right . 
\end{eqnarray}
for $\beta < \beta_c=2 \sqrt{\log 2 }/J$ in the thermodynamic limit, 
where 
$n_{\rm RS}=\beta_c^2/\beta^2>1$.  
On the other hand, 
\begin{eqnarray}
  \label{eq:conbelow}
\phi(n)=  \lim_{N\rightarrow\infty}
  \frac{1}{N}\log{\left\langle Z^n\right\rangle }
  =
  \left \{ 
    \begin{array}{ll}
      \log{2}+n^2 \beta^2 J^2/4, & n>n_c, \cr
      {n\beta J\sqrt{\log{2}}}, & n<n_c,  
    \end{array}
  \right .
\label{lowT}
\end{eqnarray}
holds for $\beta > \beta_c$, where $n_c=\beta_c/\beta<1$.

The inequality $Z\left (\{E_i\} \right )
=\sum_{i=1}^{2^N} \exp\left (-\beta E_i \right ) > 
\exp \left (-\beta E_1 \right )$ 
gives
\begin{eqnarray}
\left \langle Z^n \right \rangle^{1/N} > \left \langle e^{-n \beta E_1} 
\right \rangle^{1/N} = \exp \left (n^2 \beta^2 J^2 / 4 \right ), 
\label{carlson_REM}
\end{eqnarray}
for $n > 0$, implying that $\left \langle Z^n \right \rangle^{1/N}$ is 
not upper-bounded by $e^{\pi |n|}$. 
This means that the uniqueness of the analytical 
continuation from $n \in \mN$ to $n \in \mC$ 
(${\rm Re}(n) \ge 0$) is not guaranteed by Carlson's theorem for 
$\phi_N(n)=N^{-1} \log \left \langle Z^n \right \rangle$ 
in the continuous random energy model. 
However, RM successfully reproduces the behavior of 
eqs. (\ref{highT}) and (\ref{lowT}). 
This implies that the transitions 
$\phi(n)=\lim_{N \to \infty} \phi_N(n)$ between the multiple 
solutions that arise in RM, including the 1RSB transition, 
are not due to multiple varieties of analytical 
continuation, but are brought about by the breaking of analyticity 
with respect to the replica number $n$ in a family of REMs.

\section{Replica analysis of CDREM}
\label{replica_appendix}
We here show how $\phi_{\rm RS1}(n)$, $\phi_{\rm RS2}(n)$ and 
$\phi_{\rm 1RSB}(n)$ are derived for CDREM by the replica method (RM). 
In order to evaluate $\left \langle Z^n \right \rangle $ for 
$n \in \mR$ (or $n \in \mC$), in RM the moment is first evaluated
for $n \in \mN$, for which the power series expansion can be 
used. This yields the expression 
\begin{eqnarray}
\left \langle Z^n \right \rangle
&=&\left \langle 
\left (\sum_{A=1}^{2^N} e^{-\beta \epsilon_A} \right )^n \right \rangle 
=\sum_{A_1,A_2,\ldots,A_n} \left \langle 
e^{-\beta \sum_{a=1}^n \epsilon_{A_a} } \right \rangle  
\label{partition_expression1} \\
&=&
\sum_{
\left (p_1,p_2,\ldots,p_n \right )}
{\cal W}\left (p_1,p_2,\ldots,p_n \right )
\prod_{k=1}^n \left (I(k \beta) \right )^{p_k}, 
\label{partition_expression2}
\end{eqnarray}
where 
\begin{eqnarray}
I(\beta )=\sum_{E} P(E) e^{-\beta E} \sim \exp \left [
N\alpha \left (\log \left ( \cosh \left (\frac{\beta}{2} \right )
\right )
\right )
\right ], 
\label{I}
\end{eqnarray}
and ${\cal W}\left (p_1,p_2,\ldots,p_n \right )$ represents 
the number of ways of partitioning $n$ replicas $A_1,A_2,\ldots,A_n$
to $p_1$ states (out of $A=1,2,\ldots,2^N$) by one, to $p_2$ states by 
two, $\ldots$, and to $p_n$ states by $n$. 
Clearly, ${\cal W}\left (p_1,p_2,\ldots,p_n \right )=0$
unless $\sum_{k=1}^n k p_k =n$. 

Eq. (\ref{I}) indicates that each term in eq. (\ref{partition_expression2})
scales exponentially with respect to $N$ since ${\cal W}\left 
(p_1,p_2,\ldots,p_n \right )$ depends exponentially on 
$N$ as well. On the other hand, the number of varieties of 
partition $\left (p_1,p_2,\ldots,p_n \right )$ does not depend on $N$. 
This implies that the summation of eq. (\ref{partition_expression2})
is dominated by a single term labeled by a certain partition 
$\left (p_1^*,p_2^*,\ldots,p_n^* \right )$ 
and the exponent $\phi(n)=\lim_{N \to \infty} 
N^{-1}\log \left \langle Z^n \right \rangle$ can be accurately 
evaluated by only the single dominant term. 
Namely, $\left (p_1,p_2,\ldots,p_n \right )$ plays the role 
of a replica order parameter. It is noteworthy that this role is 
similar to that of the types in information theory. 
However, there is a distinct difference between the two notions
because $\left (p_1,p_2,\ldots,p_n \right )$ 
represents relative positions of replicas in the replicated 
system while the types are defined for a single system. 

The remaining problem is how to find the dominant partition. 
Replica symmetry, which implies that eq. 
(\ref{partition_expression1}) is invariant under any 
permutation of replica indices $a=1,2,\ldots,n$, offers a useful 
guideline for solving this problem. 
This leads to a physically natural assumption that the dominant partition is 
characterized by this symmetry as well, 
which yields the following two RS solutions:
\begin{itemize}
\item {\bf RS1:} Dominated by $\left (p_1^*,p_2^*,\ldots,p_n^* \right )
=(n,0,\ldots,0)$, giving 
\begin{eqnarray}
\left \langle Z^n \right \rangle &\sim& 2^{Nn} 
\times \left (I(\beta) \right )^n \cr
&\sim & \exp \left [N n \left (
\log 2 + 
\alpha \left (\log \left ( \cosh \left (\frac{\beta}{2} \right )
\right )
\right )
\right )
\right ]. 
\label{replica_RS1}
\end{eqnarray}
This gives $\lim_{N \to \infty}N^{-1}
\log \left \langle Z^n \right \rangle = \phi_{\rm RS1}(n)$. 

\item {\bf RS2:} Dominated by $\left (p_1^*,p_2^*,\ldots,p_n^* \right )
=(0,0,\ldots,1)$, giving 
\begin{eqnarray}
\left \langle Z^n \right \rangle &\sim& 2^{N} \times I(n \beta)   \cr
&\sim & \exp \left [N \left (
\log 2 + 
\alpha \left (\log \left ( \cosh \left (\frac{n\beta}{2} \right )
\right )
\right )
\right )
\right ]. 
\label{replica_RS2}
\end{eqnarray}
This gives $\lim_{N \to \infty}N^{-1}
\log \left \langle Z^n \right \rangle = \phi_{\rm RS2}(n)$. 
\end{itemize}

Eqs. (\ref{replica_RS1}) and (\ref{replica_RS2}) indicate
that $\phi_{\rm RS1}(n)$ and $\phi_{\rm RS2}(n)$ can be 
analytically continued from $n \in \mN$ to $n \in \mR$ (or $n \in \mC$), 
respectively. The above equations also mean that $\phi_{\rm RS1}(n)$
and $\phi_{\rm RS2}(n)$ agree at $n=1$ in general, 
which makes it difficult to choose the relevant solution for $n<1$. 
As a practical solution, we select, as the relevant
solution in this region, the solution with the larger first 
derivative, 
following an empirical criterion given in ref. \cite{OK2004}. 
This gives $\phi(n)=\phi_{\rm RS1}(n)$ for 
$\beta < \beta_c$, while $\phi(n)=\phi_{\rm RS2}(n)$ is 
chosen for $\beta > \beta_c$. However, in the latter case, 
$\phi_{\rm RS2}(n)$ is inadequate for $n<n_c=\beta_c/\beta < 1$
because the convexity condition $\left (\partial/\partial  n \right )
\left (n^{-1}\phi_{\rm RS2}(n) \right ) \ge 0$ is not satisfied. 
Therefore, we have to construct another solution 
taking the breaking of replica symmetry into account, 
which yields the 1RSB solution:
\begin{itemize}
\item {\bf 1RSB:} Dominated by $p_m^*=n/m$ for a certain $m$ and $p_k^*=0$ 
for $k \ne m$, giving 
\begin{eqnarray}
\left \langle Z^n \right \rangle &\sim& 2^{Nn/m} \times \left (I(m \beta) 
\right )^{n/m}  \cr
&\sim & \exp \left [N \frac{n}{m}\left (
\log 2 + 
\alpha \left (\log \left ( \cosh \left (\frac{m\beta}{2} \right )
\right )
\right )
\right )
\right ]. 
\label{replica_1RSB}
\end{eqnarray}
After analytical continuation, $m$ is 
determined so as to extremize the right hand side, which 
leads to $m=n_c=\beta_c/\beta$. 
It may be noteworthy that the extremum means
not maximum but minimum. 
This gives $\lim_{N \to \infty}N^{-1}
\log \left \langle Z^n \right \rangle = \phi_{\rm 1RSB}(n)$. 
\end{itemize}
Figure \ref{phi_CDREM} schematically shows the profiles of the three solutions.
%



\section*{References}

\begin{thebibliography}{99}
\bibitem{Beyond1987} M\'{e}zard M, Parisi G and Virasoro M A
1987 {\em Spin Glass Theory and Beyond} (Singapore: World Scientific)
\bibitem{Dotzenko2001} 
Dotzenko V S 2001 {\em Introduction to the Replica Theory of Disordered Statistical 
Systems} (Cambridge: Cambridge University Press)
\bibitem{Talagrand2003}
Talagrand M 2003 
{\em Spin Glasses: A Challenge for Mathematicians. Mean Field Models and Cavity Method}
(Berlin: Springer Verlag)
\bibitem{Talagrand2006} 
Talagrand M 2006 {\em Ann. Math.} {\bf 163} 221 
\bibitem{Titchmarsh1939} 
Titchmarsh E C 1932 {\em The Theory of Functions} 
(Oxford: Oxford University Press)
\bibitem{SK1975}
Sherrington D and Kirkpatrick S
1975 {\em Phys. Rev. Lett.} {\bf 35} 1972 
\bibitem{vanHemmen1978} 
van Hemmen J L and Palmer R G 1979 \JPA {\bf 12} 563
\bibitem{Mou1} Moukarzel C and Parga N 1991 {\em Physica } 
{\bf A177} 24
\bibitem{Mou2} Moukarzel C and Parga N 1992 {\em Physica}
{\bf A185} 305
\bibitem{OK2004} Ogure K and Kabashima Y 2004 
{\em Prog. Theor. Phys.} {\bf 111} 661
\bibitem{Derrida1981} Derrida B 1981 {\em Phys. Rev. B} {\bf 24} 2613
\bibitem{GardnerDerrida1989}
Gardner E and Derrida B 1989 \JPA {\bf 22} 1975
\bibitem{CsiszarKorner1981}
Csisz\'{a}r I and K\"{o}rner J 
1982 {\em Information Theory: Coding Theorems for Discrete Memoryless Systems}
(Orlando: Academic Press)
\bibitem{BouchaudMezard1997}
Bouchaud J-P and M\'{e}zard M 1997  \JPA {\bf 30} 7997
\bibitem{Gumbel1958}
Gumbel E J 1958 {\em Statistics of Extremes} (New York: Columbia University Press)
\bibitem{LeeYang}
Lee L D and Yang C N 1952 {\em Phys. Rev. } {\bf 87} 410 
\bibitem{Fisher} Fisher M E 1965 {\em Lectures in Theoretical Physics } vol. 7
(Boulder: University of Colorado Press)
\bibitem{OK2005} Ogure K and Kabashima Y 2005 
{\em Prog. Theor. Phys. Suppl.} {\bf 157} 103
\bibitem{Obuchi2009} Obuchi Y, Kabashima Y and Nishimori H
2009 {\em J. Phys. A: Math. Theor.} {\bf 42} 075004 (27pp)
\bibitem{Derrida1991} Derrida B 1991 {\rm Physica } {\bf A177} 31
\bibitem{OK2008b} Ogure K and Kabashima Y {\em  
On analyticity with respect to the replica number in 
random energy models II: zeros on the complex plane}
{\em Preprint} arXiv:0812.4655 



































\end{thebibliography}
\end{document}